\documentclass{ws-procs9x6}

\begin{document}

\title{
 Nuclear Superfluidity in Exotic Nuclei and Neutron Stars}

\author{
Nicolae Sandulescu}

\address {
Institute of Physics and Nuclear Engineering, 76900 Bucharest, 
  Romania}

\begin{abstract}
 Nuclear superfludity in exotic nuclei close to the drip lines and
 in the inner crust matter of neutron stars have common features
 which can be treated with the same theoretical tools. In the
 first part of my lecture I discuss how two such tools,
 namely the HFB approach and the linear response theory can be
 used to describe the pairing correlations in weakly
 bound nuclei, in which the unbound part of the energy spectrum
 becomes important. Then, using the same models, I shall discuss
 how the nuclear superfluidity can affect the thermal properties of
 the inner crust of neutron stars.

\end{abstract}

\bodymatter
%%%%%%%%%%%%%%%%%%%%%%%%%%%%%%%%%%%%%%%%%%%%%%%%%%%%%%%%%%%
\section{Introduction}
%%%%%%%%%%%%%%%%%%%%%%%%%%%%%%%%%%%%%%%%%%%%

 The basic features of nuclear superfluidity are the same in finite nuclei and
 in infinite Fermi systems such as neutron stars. Yet, in atomic nuclei the pairing 
 correlations have special features related to the finite size of the system.
 The way how the finite size affects
 the pairing correlations depends on the position of the  chemical potential. If the
 chemical potential is deeply bound, like in stable and heavy nuclei, the finite size
 influences the  pairing correlations mainly  through the shell structure induced by
 the spin-orbit interaction. The situation becomes more complex in weakly bound nuclei
 close to the drip lines, where the chemical potential approaches the continuum threshold.
 In this case the inhomogeneity of the pairing field can produce a strong
 coupling between the bound and  the unbound part of the single-particle spectrum.
 This is an issue which will be discussed in the first part of my lecture. More
 precisely, I shall discuss how the continuum coupling and the pairing correlations
 can be treated in the framework of the Hartree-Fock-Bogoliubov (HFB) approach and
  linear response theory.
 
 The neutron-drip line put a limit to the  neutron-rich nuclei which can be 
 produced in the laboratory or in supernova explosions (via rapid neutron-cupture
 processes). However, this limit can be overpassed in the inner crust of neutron stars
 since there the driped neutrons are kept together with the neutron-rich nuclei by the
 gravitational preasure.

 The superfluid properties of the inner crust have been 
 considered long ago in connection with post-glitch timing
 observations and colling processes \cite{pines,saul,prakash}. However, 
 although the  neutron star matter  superfluidity has been intensively 
 studied in the last decades \cite{lombardo}, so far only a  few 
 microscopic calculations have been done for the  superfluidity of inner 
 crust matter. The most sophisticated microscopic calculations done till now 
  use  the  framework of HFB approach
  at finite temperature \cite{ns2}. They will be discussed in the second
  part of my lecture. The discussion  will be focused on the effects induced by
  the pairing correlations on the specific heat and on the cooling time of the
  inner curst of neutron stars. 
 
\section{Pairing correlations in exotic nuclei}

\subsection{Continuum-HFB approach}

 The tool commonly used for treating  pairing correlations in exotic nuclei
 close to the drip lines is the Hartree-Fock-Bogoliubov (HFB) approach \cite{rs}.
 The novel feature of interest here is how within this approach one can 
 treat properly the quasiparticle states belonging to the continuum
 spectrum \cite{chfb}. This issue is discussed below.

 The HFB equations for local fields and spherical symmetry have the form:
 \begin{equation}
\begin{array}{c}
\left( \begin{array}{cc}
h(r) - \lambda & \Delta (r) \\
\Delta (r) & -h(r) + \lambda
\end{array} \right)
\left( \begin{array}{c} U_i (r) \\
 V_i (r) \end{array} \right) = E_i
\left( \begin{array}{c} U_i (r) \\
 V_i (r) \end{array} \right) ~,
\end{array}
\label{1}
\end{equation}
\\
where $\lambda$ is the chemical potential, $h(r)$ is the mean field hamiltonian
and  $\Delta(r)$ is the pairing field. The fields depend on particle density $\rho(r)$ 
and pairing density $\kappa(r)$ given by:
\begin{equation}
\rho(r) =\frac{1}{4\pi} \sum_{i} (2j_i+1) V_i^* (r)
V_i (r) ,
\label{2}
\end{equation}
\begin{equation}
\kappa(r) = \frac{1}{4\pi} \sum_{i} (2j_i+1) U_i^* (r)
V_i (r) .
\label{3}
\end{equation}
In the calculations presented here the mean field is described with
a Skyrme force and for the pairing interaction it is used a density- dependent
force of zero range of the following form \cite{bertsch}:
\begin{equation}
V (\mathbf{r}-\mathbf{r^\prime}) = V_0 [1 -
\eta(\frac{\rho}{\rho_0})^{\alpha}]
\delta(\mathbf{r}-\mathbf{r^\prime})
\equiv V_{eff}(\rho(r)) \delta(\mathbf{r}-\mathbf{r^\prime}).
\end{equation}
For this force the pairing field is given by
$\Delta(r) = \frac{V_{eff}}{2} \kappa (r)$.

The HFB equations have two kind of solutions. Thus, between 0 and $-\lambda$ 
the quasiparticle spectrum is discrete and both upper and lower components of
the radial HFB wave function decay exponentially at infinity. On the other hand,
for $E > - \lambda$ the spectrum is continuous and the solutions are:
\begin{eqnarray}
u_{lj}(E,r) & = & C [cos(\delta_{lj}) j_l(\alpha_1r)-sin(\delta_{lj})
n_l(\alpha_1r)]~, \nonumber \\
v_{lj}(E,r) & = & D_1 h_l^{(+)}(i\beta_1r)~,
\label{eq3}
\end{eqnarray}
where $j_l$ and $n_l$ are spherical Bessel and Neumann functions repectively and $\delta_{lj}$
is the phase shift corresponding to the angular momentum $(lj)$.
The phase shift is found by matching the asymptotic form of the wave function written
above with the inner radial wave function. From the energy dependence of the
phase shift one can determine the energy regions of quasiparticle
resonant states. In HFB they are of two types. A first type corresponds to
the single-particle resonances of the mean field. A second kind of resonant 
states is specific to the HFB approach and
corresponds to the bound single-particle states which in the absence of pairing
correlations have an energy $\epsilon < 2\lambda$. Among all possible resonances,
of physical interest are the ones close to the continuum threshold. An exemple of
such resonances is shown in section 2.4.

\subsection{Resonant states in the BCS approach}

The low-lying quasiparticle resonances can be also calculated in the framework
 of the BCS approach. The BCS approximation is obtained by 
 neglecting in the HFB equations the non-diagonal matrix elements of
 the pairing field. This means that  in the BCS limit one neglects the pairing correlations
 induced by the pairs formed in states which are not time-reversed
 partners.

 The extension of BCS equations for taking into account the 
  effect of resonant states was proposed in Refs.\cite{hfbcs,bcs}.
 For the case of a general pairing interaction the corresponding
 resonant-BCS equations read \cite{hfbcs}:
 
\begin{equation}\label{eq:gapr1}
\Delta_i = \sum_{j}V_{i\overline{i}j\overline{j}} u_j v_j +
\sum_\nu
V_{i\overline{i},\nu\epsilon_\nu\overline{\nu\epsilon_\nu}}
\int_{I_\nu} g_{\nu}(\epsilon)
u_\nu(\epsilon) v_\nu(\epsilon) d\epsilon~,
\end{equation}
\begin{equation}\label{eq:gapr}
\Delta_\nu \equiv
\sum_{j}
V_{\nu\epsilon_\nu\overline{\nu\epsilon_\nu},j\overline{j}} u_j v_j +
\sum_{\nu^\prime}
V_{\nu\epsilon_\nu\overline{\nu\epsilon_\nu},
\nu^\prime\epsilon_{\nu^\prime}
\overline{\nu^\prime\epsilon_{\nu^\prime}}}
\int_{I_{\nu^\prime}} g_{\nu^\prime}(\epsilon^\prime)
u_{\nu^\prime}(\epsilon^\prime) v_{\nu^\prime}(\epsilon^\prime)
d\epsilon^\prime~,
\end{equation}
\begin{eqnarray}
N = \sum_i v_i^2 + \sum_\nu \int_{I_\nu} g_{\nu}(\epsilon) v^2_\nu
(\epsilon) d\epsilon~.
\label{eq15}
\end{eqnarray}
Here $\Delta_i$ is the gap for the bound state $i$ and
$\Delta_\nu$ is the averaged gap for the resonant state $\nu$.
The quantity $g_\nu(\epsilon) = \frac {2j_\nu +1}{\pi}
\frac{d\delta_\nu}{d\epsilon}$ is the continuum level density 
and
$\delta_\nu$ is the phase shift of angular momentum $(l_{\nu} j_{\nu})$.
The factor $g_\nu(\epsilon)$ takes into account the variation
of the localisation of scattering states in the energy region of
a resonance ( i.e., the width effect) and becomes a delta function in the
limit of a very narrow width. 
In these equations the interaction matrix elements are  calculated with
the scattering wave functions at resonance energies and normalised
inside the volume where the pairing interaction is active. 

The BCS equations written above have been solved with a single particle spectrum
 corresponding to a HF \cite{hfbcs} and a RMF \cite{rmfbcs} mean field.
 It was shown that by including
 in the BCS equations a few relevant resonances close to the continuum threshold
 one can get results very similar to the ones obtained in the HFB and  RHB
 calculations. One can thus conclude that in nuclei close to the dripline
 the quasiparticle spectrum is dominated by a few low-lying resonances.
 In many odd-even nuclei close to the drip lines these low-lying quasiparticle
 resonances might be the only measurable excited states. Their widths
 can be calculated from the phase shift behaviour or from the imaginary part
 of the energies associated to the Gamow states \cite{gamowbcs}. 

 In even-even nuclei the low-lying quasiparticle resonances could form the main
 component of unbound collective excitations with a finite time life. How these
 excitations can be treated in nuclei close to the drip line is discussed in
 the next section.
  
\subsection{Linear response theory with pair correlations and 
continuum coupling}

 The collective excitations of atomic nuclei in the presence of pairing
 correlations is usually described in the Quasiparticle-Random Phase
 Approximation (QRPA) \cite{rs}. In nuclei characterized by a
 small nucleon separation energy, the excited states are strongly influenced
 by the coupling with the quasiparticle continuum configurations. Among
 the configurations of particular interest are the two-quasiparticle states in which
 one or both quasiparticles are in the continuum. In order to describe such
 excited states one needs a proper treatment of the continuum coupling, which
 is missing in the usual QRPA calculations based on a
 discrete quasiparticle spectrum. In this section we discuss how  the pairing and
 the continuum coupling can be treated in the framework of the linear response 
 theory \cite{cqrpa,transfer}. 

 The response of the nuclear system to an external perturbation is obtained from the
 time-dependent HFB equations (TDHFB) \cite{rs}:
\begin{equation}\label{eq:tdhfb}
i\hbar\frac{\partial{\cal R}}{\partial t}=[{\cal H}(t) +
{\cal F}(t),{\cal R}(t)],
\end{equation}
where ${\cal R}$ and ${\cal H}$ are the time-dependent generalized density and
the HFB Hamiltonian, respectively. ${\cal F}$ is the external oscillating field :
\begin{equation}\label{eq:pert}
{\cal F} = F e^{-i\omega t} + h.c. .
\end{equation}
In Eq. (\ref{eq:pert}) $F$ includes both particle-hole and two-particle transfer
operators :
\begin{equation}\label{eq:extpart}
F=\sum_{ij} F^{11}_{ij} c_{i}^{\dagger}c_{j}+\sum_{ij}
(F^{12}_{ij} c_{i}^{\dagger}c_{j}^{\dagger}+ F^{21}_{ij} c_{i}c_{j}),
\end{equation}
and $c_{i}^{\dagger}$, $c_{i}$ are the particle creation and annihilation
operators, respectively.

In the small amplitude limit the TDHFB equations become:
\begin{equation}\label{eq:lin}
        \hbar\omega{\cal R}'=[{\cal H}',{\cal R}^0] + [{\cal H}^0,{\cal
	        R}']+[F,{\cal R}^0],
\end{equation}
where the superscript \ ' stands for the corresponding perturbed quantity. 

The variation of the generalized density ${\cal R}$' is expressed in
term of 3 quantities, namely $\rho'$, $\kappa'$ and $\bar{\kappa}'$,
which are written as a column vector : 

\begin{equation}\label{eq:rhodef}
\rho'       =\left(
	\begin{array}{c}
	\rho' \\
         \kappa' \\
	 \bar{\kappa}' \\
        \end{array}
	\right),
	\end{equation}
where $\rho'_{ij} = \left<0|c^{\dagger}_jc_i|'\right>$
is the variation of the particle density,
$\kappa'_{ij} =\left<0|c_jc_i|'\right>$ and $\bar{\kappa}'_{ij} =
\left<0|c^{\dagger}_jc^{\dagger}_i|'\right>$ are the
fluctuations of the pairing tensor associated to the pairing vibrations and
$\mid ' \rangle$ denotes the change of the ground state wavefunction $|0>$
due to the external field. 

The variation of the HFB Hamiltonian is given by:

\begin{equation}\label{eq:hvar}
{\cal H}'=	\bf{V}\rho',
\end{equation}
where $\bf{V}$ is the matrix of the residual interaction 
expressed in terms of the second derivatives of the HFB energy 
functional, namely:
					   
\begin{equation}\label{eq:vres}
{\bf{V}}^{\alpha\beta}({\bf r}\sigma,{\bf r}'{\sigma}')=
\frac{\partial^2{\cal E}}{\partial{\bf{\rho}}_\beta({\bf r}'{\sigma}')
\partial{\bf{\rho}}_{\bar{\alpha}}({\bf r}\sigma)},~~~\alpha,\beta = 1,2,3.
\end{equation}
In the above equation the notation $\bar{\alpha}$ means that whenever $\alpha$ is 2 or 3
then $\bar{\alpha}$ is 3 or 2.

Introducing for the external field the three dimensional column vector:

\begin{equation}\label{eq:f}
\bf{{F}}=	\left(\begin{array}{c}
	 {F}^{11} \\
	 {F}^{12}\\
	 {F}^{21}   \\
	\end{array}
	\right),
\end{equation}
the density changes can be written in the standard form:

\begin{equation}\label{eq:g}
{\rho'}=\bf{G}\bf{F}~\boldmath \everymath{ },
\end{equation}
where $\bf{G}$ is the QRPA Green's function obeying the Bethe-Salpeter equation:

\begin{equation}\label{eq:bs}
\bf{G}=\left(1-\bf{G}_0\bf{V}\right)^{-1}\bf{G}_0=\bf{G}_0+\bf{G}_0\bf{V}\bf{G}.
\end{equation}
The unperturbed Green's function $\bf{G}_0$ has the form:

\begin{equation}\label{eq:g0}
{\bf{G}_0}^{\alpha\beta}({\bf r}\sigma,{\bf r}'{\sigma}';\omega)=
\sum_{ij} \hspace{-0.5cm}\int \hspace{0.3cm} \frac{{\cal U}^{\alpha 1}_{ij}({\bf r}\sigma)
\bar{{\cal U}}^{*\beta 1}_{ij}({\bf r}'\sigma')}{\hbar\omega-(E_i+E_j)+i\eta}
-\frac{{\cal U}^{\alpha 2}_{ij}({\bf r}\sigma)
\bar{{\cal U}}^{*\beta 2}_{ij}({\bf r}'\sigma')}{\hbar\omega+(E_i+E_j)+i\eta},
\end{equation}
where $E_i$ are the qp energies and ${\cal U}_{ij}$ are 3 by 2
matrices expressed in term of the two components of the HFB wave functions \cite{cqrpa}. 
The $\sum$ \hspace{-0.45cm}$\int$ ~ symbol in Eq. (\ref{eq:g0}) indicates that the
summation is taken over the discrete and the continuum quasiparticle  states.

The QRPA Green's function can be used for calculating the strength function
associated with various external perturbations. For instance, the transitions from the
ground state to the excited states induced by a particle-hole external field
can be described by the strength function:

\begin{equation}\label{eq:stren}
S(\omega)=-\frac{1}{\pi}Im \int  F^{11*}({\bf r}){\bf{
G}}^{11}({\bf r},{\bf r}';\omega) F^{11}({\bf r}')
d{\bf r}~d{\bf r}'
\end{equation}
where ${\bf{G}}^{11}$ is the (ph,ph) component of the QRPA Green's function.
Another process which can be described in the same manner is the two-particle
transfer from the ground state of a nucleus with A nucleons to the  excited states of
a nucleus with A+2 nucleons. For this process the strength function is: 

\begin{equation}\label{eq:stren2}
S(\omega)=-\frac{1}{\pi}Im \int  F^{12*}({\bf r}){\bf{
G}}^{22}({\bf r},{\bf r}';\omega) F^{12}({\bf r}')
d{\bf r}~d{\bf r}'
\end{equation}
where ${\bf{G}}^{22}$ denotes the (pp,pp) component of the 
Green's function. 
 
\subsection{Quasiparticle excitations and two-neutron transfer in neutron-rich nuclei}

 The formalisms presented above are illustrated here for the case  of neutron-rich 
 oxygen isotopes \cite{cqrpa,transfer}. First, we present an exemple of quasiparticle
 resonances calculated in the framework of the continuum-HFB (cHFB) approach introduced
 in section 3.1.  In the cHFB calculations 
 the mean field quantities are evaluated using the Skyrme 
 interaction SLy4 \cite{sly4}, while for the pairing interaction we take a 
 zero-range density-dependent force. The parameters of the pairing force are 
 given in Ref \cite{cqrpa}.  The HF single-particle and HFB quasiparticle energies
 corresponding to the $sd$ shell and to the $1f_{7/2}$ state are listed in 
 Table 1. One can notice that in both HF and cHFB calculations the 
 state $1f_{7/2}$ is a wide resonance for $^{18-22}$O nuclei, while the
 state $1d_{3/2}$ is a narrow resonance. As seen below, these one-quasiparticle
 resonances determine essentially the two-quasiparticle states which are the stongest
 populated in even-even oxygen isotopes. \\
 
\noindent Table 1. HF and HFB energies in oxygen isotopes.
\begin{table}[h]
\begin{tabular}{|c|c||c|c|c|}
\hline
&     & $^{18}$O   & $^{20}$O  & $^{22}$O
\\ \hline
1d$_{5/2}$ & HF & -6.7 & -7.0 & -7.45
\\ \hline
1d$_{5/2}$ & cHFB & 2.26 & 2.08 & 2.30
\\ \hline
2s$_{1/2}$ & HF & -4.0 & -4.2 & -4.6
\\ \hline
2s$_{1/2}$ & cHFB & 3.46 & 2.28 & 1.05
\\ \hline
1d$_{3/2}$ & HF & (0.46;0.02) & (0.51;0.03) & (0.42;0.02)
\\ \hline
1d$_{3/2}$ & cHFB & (7.74;0.12)  & (6.60;0.29) & (5.39;0.01)
\\ \hline
1f$_{7/2}$ & HF & (5.50;1.35) & (5.24;1.24)  &  (4.86;1.04)
\\ \hline
1f$_{7/2}$ & cHFB &(12.77;1.13)  & (12.14;0.83) & (10.05;0.69)
\\ \hline
\end{tabular}
\end{table}

 The two-quasiparticle states are calculated by using the response theory
 described in section 2.3. In the calculations one includes the full
 discrete and continuum HFB spectrum up to 50 MeV. These states are used to
 construct the unperturbed Green's function ${\bf G_{0}}$. After solving the
 Bethe-Salpeter equation for the QRPA Green function one constructs the strength
 functions written in section 2.3. The details of the calculations can be found in 
 Ref\cite{transfer}.
 Here we show the results obtained for the collective states excited in the 
 two-neutron transfer. The strength function corresponding to a neutron pair
transferred to the oxygen isotope $^{22}$O is shown in Fig.1. 

\begin{figure}[h] 
\begin{center}
\includegraphics*[scale=0.35]{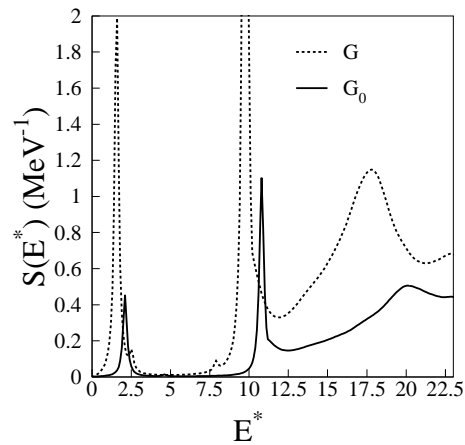}
\caption{The response function for the two-neutron transfer on
22O. The unperturbed response is in solid line and the QRPA response in dashed line.} 
\end{center}
\end{figure}

For the isotope $^{22}$O the subshell $d_{5/2}$ is essentially blocked for the pair
transfer. Therefore in this nucleus we can clearly identify only two peaks
below 11 MeV, corresponding to a pair transferred to the states $2s_{1/2}$
and $2d_{3/2}$. The strength function shown in Fig. 1 shows also a broad
 peak around 20 MeV. This peak is built mainly upon
the single-particle resonance $1f_{7/2}$ and its cross section is much
larger than the one associated to the lower energy transfer modes. Since
this high energy transfer mode is formed mainly by single-particle states
above the valence shell, this mode is similar to the giant pairing
vibration mode suggested long ago \cite{liotta}. Although such a mode has not been
detected yet, the pair transfer reactions involving exotic loosely bound
nuclei may offer a better chance for this undertaking \cite{fortunato}.

\section{Superfluid and thermal properties of neutron stars crust}
  
 In this section we analyse the superfluid properties of a nuclear
 system in which the limit of the neutron-drip line is overpassed:
 the inner crust of neutron stars. The inner crust consists of a 
 lattice of neutron-rich nuclei immersed in a sea of unbound neutrons
 and relativistic electrons. Down to the inner edge of the
 crust, the crystal lattice is most probably formed by 
 spherical nuclei. More inside the star, before the 
 nuclei dissolve completely into the liquid of the core, 
 the nuclear matter can develop other exotic configurations
 as well, i.e, rods,  plates, tubes, and bubbles \cite{pethick}.
 The thickness of the inner crust is rather small, of the order
 of one kilometer, and its mass is only about  1$\%$ of the 
 neutron star mass. However, in spite of its small size, the properties
 of inner crust matter, especially its superfluid  properties, 
 have important consequences for the dynamics  and the thermodynamics
 of neutron stars. In what follows we will discuss the main features of the
  pairing correlations inside the inner crust matter. Then we shall focus
  on the effects induced by the nuclear superfluidity on the specific
  heat and on the cooling time of the curst.

\subsection{Superfluid properties of the inner crust matter}
 
 The superfluid properties of the inner crust matter discussed here
 \cite{ns2} are based on the finite-temperature HFB (FT-HFB) approach.
 For zero range forces and spherical symmetry the  radial FT-HFB have
 a similar form as the HFB equations at zero temperature, ie.,
\begin{equation}
\begin{array}{c}
\left( \begin{array}{cc}
h_T(r) - \lambda & \Delta_T (r) \\
\Delta_T (r) & -h_T(r) + \lambda 
\end{array} \right)
\left( \begin{array}{c} U_i (r) \\
 V_i (r) \end{array} \right) = E_i
\left( \begin{array}{c} U_i (r) \\
 V_i (r) \end{array} \right) ~,
\end{array}
\label{1}
\end{equation}
where $E_i$ is the quasiparticle energy, $U_i$, $V_i$ are 
the components of the radial FT-HFB wave function
and $\lambda$ is the chemical potential.
The quantity $h_T(r)$ is the thermal averaged mean field 
hamiltonian and $\Delta_T (r)$ is the thermal averaged
pairing field. The latter depends on the average pairing
density $\kappa_T$. In a self-consistent calculation based on a 
Skyrme-type force , $h_T(r)$ is expressed in terms of thermal
averaged densities, i.e., kinetic energy density $\tau_T(r)$, particle 
density $\rho_T(r)$ and spin density $J_T(r)$, in the same way as in 
the Skyrme-HF approach. The thermal averaged densities
mentioned above are given by \cite{ns2}:
\begin{equation}
\rho_T(r) =\frac{1}{4\pi} \sum_{i} (2j_i+1) [ V_i^* (r) 
V_i (r) (1 - f_i ) \\
+ U_i^* (r) U_i (r) f_i ] 
\nonumber
\end{equation}
\begin{equation}
\kappa_T(r)=\frac{1}{4\pi} \sum_{i} (2j_i+1) U_i^* (r) V_i (r)
(1 - 2f_i ) \nonumber
\end{equation}
\begin{eqnarray}
J_T(r) & = & \frac{1}{4\pi} \sum_i (2j_i+1) 
[j_i(j_i+1)-l_i(l_i+1)-\frac{3}{4}] \nonumber  \\
& & \times \{ V_i^2 (1-f_i) + U_i^2 f_i \}
\nonumber
\end{eqnarray}
\begin{eqnarray}
\tau_T(r) & = & \frac{1}{4\pi} \sum_{i} (2j_i+1)
\{ [(\frac{dV_i}{dr}-\frac{V_i}{r})^2 +\frac{l_i(l_i+1)}{r^2} V_i^2 ]
\nonumber \\
& & \times (1 - f_i)
+ [(\frac{dU_i}{dr}-\frac{U_i}{r})^2 +\frac{l_i(l_i+1)}{r^2}
U_i^2]f_i \} , \nonumber \\
\end{eqnarray}
where $f_i = [1 + exp ( E_i/k_B T)]^{-1}$ is the 
Fermi distribution, $k_B$ is the Boltzmann constant and T is the 
temperature. The summations in the equations above are over the
whole quasiparticle spectrum, including the unbound states. 

The FT-HFB equations are solved for the spherical
Wigner-Seitz  cells determined in Ref.\cite{negele}.
To generate far from the nucleus a constant density
corresponding to the neutron gas, the FT-HFB equations are 
solved by imposing Dirichlet-Neumann boundary conditions
at the edge of the cell \cite{negele}, i.e., all wave
functions of even parity vanish and the derivatives of 
odd-parity wave functions vanish.
 In the FT-HFB calculations we use for the particle-hole channel
 the Skyrme effective interaction SLy4 \cite{sly4}, which 
 has been adjusted to describe properly the mean field properties
 of neutron-rich nuclei and infinite neutron matter. 
 In the particle-particle channel we employ a density dependent zero range force.
 Since the magnitude of pairing
 correlations  in neutron matter is still a subject of debate, the 
 parameter of the pairing force are chosen so as to describe two different
 scenarios for the neutron matter superfluidity. 
 Thus,  for the first calculation we use the parameters: $V_{0}$=-430 MeV  fm$^3$, 
 $\eta$=0.7, and $\alpha$=0.45. With these parameters and
 with a  cut-off energy for the quasiparticle spectrum equal
 to 60 MeV one obtains approximately the pairing gap
 given by the Gogny force in nuclear matter \cite{bertsch}.
 In the second calculation we reduce the strength of the force
 to the value $V_0$=-330 MeV fm$^3$. With this value of the 
 strength we simulate the second scenario for the neutron
 matter superfluidity, in which the screening effects would reduce 
 the maximum gap in neutron matter to a value around 1 MeV \cite{screening}.

 The FT-HFB results are shown here for two representative 
 Wigner-Seitz cells chosen from Ref.\cite{negele}.
 These cells contain Z=50 protons and have rather different
 baryonic densities, i.e., 0.0204 fm$^{-3}$ and  0.00373 fm$^{-3}$. 
 The cells, which contain N=1750 and N=900 neutrons, respectively,
 are denoted below as a nucleus with Z protons and N neutrons,
 i.e., $^{1800}$Sn and $^{950}$Sn. The FT-HFB calculations are done
 up to a maximum temperature of T=0.5 MeV, which is covering the
 temperature range of physical interest \cite{riper}.
 
\begin{figure}[h]
\begin{center}
\includegraphics*[scale=0.35,angle=-90]{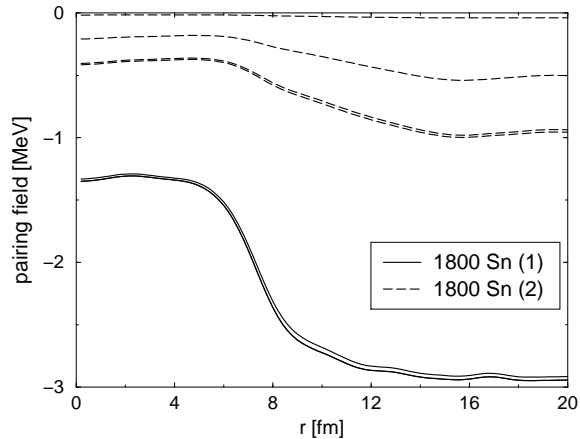}
\caption{
Neutron pairing fields for the cell $^{1800}$Sn
calculated at various temperatures. The numbers 1 and 2 
which follow the cell symbol (see the inset) indicate 
the variant of the pairing force used in the calculations. 
The full and the dashed lines corresponds (from bottom upwards)
to the set of temperatures T=$\{0.0, 0.5\}$
MeV and T=$\{0.0, 0.1, 0.3, 0.5\}$MeV, respectively.}
\end{center}
\end{figure}

 The temperature dependence of the pairing fields in the two cells
 presented above is shown in Figs.2-3. First, we notice that for
 all temperatures the nuclear clusters modify significantly the
 profile of the pairing field. One can also see that for most of
 the cases the temperature  dependence of the pairing field is
 significant. This is clearly seen in the low-density cell $^{950}$Sn.
 
\begin{figure}[h]
\begin{center}
\includegraphics*[scale=0.35,angle=-90]{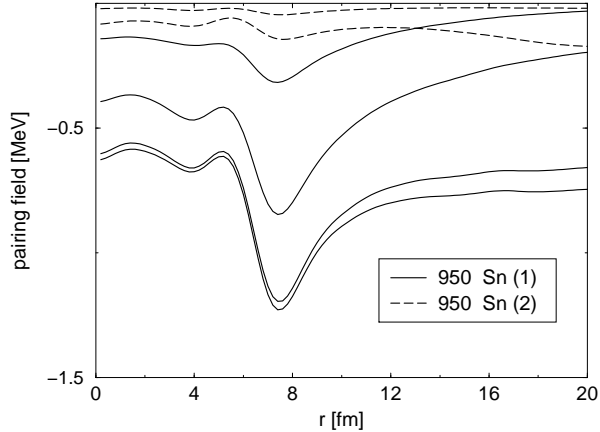}
\caption{The same as in Fig.1, but for the cell $^{950}$Sn.
The full and the dashed lines corresponds (from bottom 
upwards) to the set of temperatures T=$\{0.0, 0.1, 0.3, 0.5\}$
and T=$\{0.0, 0.1\}$MeV, respectively}
\end{center}
\end{figure}

\subsection{Specific heat of the inner crust baryonic matter}

The superfluid properties of the neutrons discussed in the previous section
have a strong influence on the specific heat of the inner crust matter \cite{ns2}.
The specific heat of a given cell of volume V is defined by:

\begin{equation}
C_V = \frac{1}{V}\frac{\partial \mathcal{E}(T)}{\partial T} ,
\end{equation}
where $\mathcal{E}(T)$ is the total energy 
of the baryonic matter inside the cell, i.e.,

\begin{equation}
\mathcal{E}(T) = \sum_i f_i E_i .
\end{equation}
 Due to the energy gap in the excitation spectrum, the specific
 heat of a superfluid system is dramatically  reduced  compared
 to its value in the normal phase. Since the specific heat depends
 exponentially on the energy gap, its value for a Wigner-Seitz cell
 is very sensitive to the local variations of the pairing field
 induced by the nuclear clusters. This can be clearly
 seen in Fig.4, where the specific heat is plotted for the cell 
 $^{1800}$Sn and for the neutrons uniformly distributed
 in the same cell. One can notice that at T=0.1 MeV and for
 the first pairing force the presence of the cluster increases
 the specific heat by about 6 times  compared to the value for
 the uniform neutron gas.
However, the most striking fact seen in Fig.4 is the huge difference
 between the predictions of the two pairing forces. Thus,
 for T=0.1 MeV this difference amounts to about 7 orders
 of magnitude. 

\begin{figure}[h]
\begin{center}
\includegraphics*[scale=0.35,angle=-90.]{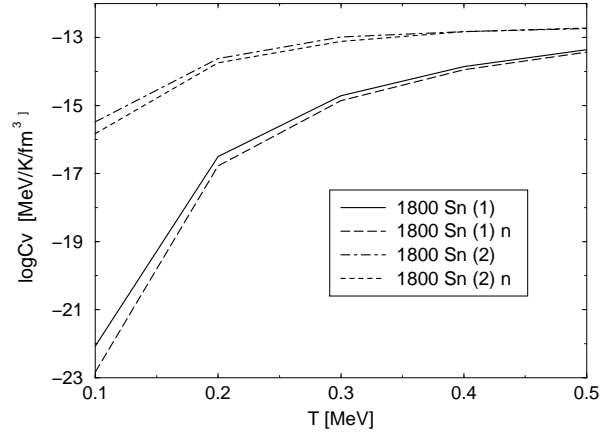}
\caption{
Specific heat for the cell $^{1800}$Sn as a function of
temperature. The notations used in the inset and the
representation of the calculated values are the same
as in Figs.1-3.}
\end{center}
\end{figure}

\begin{figure}[h]
\begin{center}
\includegraphics*[scale=0.35,angle=-90.]{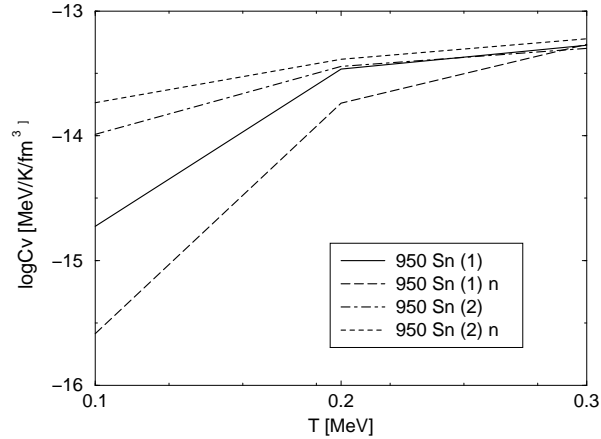}
\caption{ The same as in Fig.4, but for the cell $^{950}$Sn.}
\end{center}
\end{figure}

 The behaviour of the specific heat for the low- density cell
 $^{950}$Sn is shown in Fig.5. For the first pairing force
 we can also see that at T=0.1 MeV the cluster increases
 the specific heat by about the same factor as in the
 cell $^{1800}$Sn. However, for the second pairing force
 the situation is opposite: the presence of the nucleus 
 decreases the specific heat instead of increasing it. 

\subsection{Collective modes in the inner crust matter}

 In the calculations presented in the previous section the specific heat of inner
 crust matter was evaluated by considering only non-interacting 
 quasiparticles states. However, the specific heat can be also strongly
 affected by the collective modes created by the residual interaction
 between the quasiparticles, especially if these modes appear at
 low-excitation energies.  

 The collective modes in the inner crust matter were calculated
 in Ref. \cite{ns3} in the framework of linear response theory discussed
 in section 2.3. The most important result of these calculations is the apparence of
 very collective modes at low energies, of the order of the pairing gap.
 An example of such mode is seen in Fig.6, were is shown  the quadrupole
 response for the cell $^{1800}$Sn. As can be clearly seen from Fig.6 ,
 when the residual interaction is introduced among the quasiparticles
 the unperturbed spectrum, distributed over a large energy region, is
 gathered almost entirely in the peak located at about 3 MeV.
 This peak collects more than 99$\%$ of the total quadrupole
 strength and is extremely collective. An indication of
 the extreme collectivity of this low-energy mode can be also seen from its
 reduced transition probability, B(E2), which is equal to 25.10$^3$ Weisskopf units.
 This value of B(E2) is two orders of magnitude higher than in standard nuclei.
 This underlines the fact that this WS cell cannot be simply considered as a giant
 nucleus. The reason is that in this cell the collective dynamics of the neutron
 gas dominates over the cluster contribution.

\begin{figure}[h]
\begin{center}
\includegraphics*[scale=0.35]{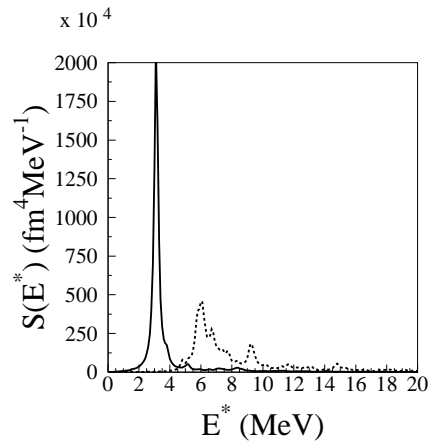}
\caption{ Quadrupole strength distribution of neutrons for the cell
$^{1800}$Sn. The full curve represents the QRPA strength, and the dashed
line is the HFB unperturbed strength.}
\end{center}
\end{figure}

  The collective excitations located at low energies can affect significantly the
  specific heat of the inner crust baryonic matter. This can be seen in Fig.7,
  where the specific heat corresponding to the collective modes (of multipolarity
   L=0,1,2,3) is shown. We notice that for T=0.1 MeV the specific heat given by
  the collective modes is of the same order of magnitute  as the one corresponding
  to HFB spectrum. Therefore one expects that the collective modes of the inner crust
  matter could affect strongly the thermal behaviour of the crust.

\begin{figure}[h]
\begin{center}
\includegraphics*[scale=0.35,angle=-90.]{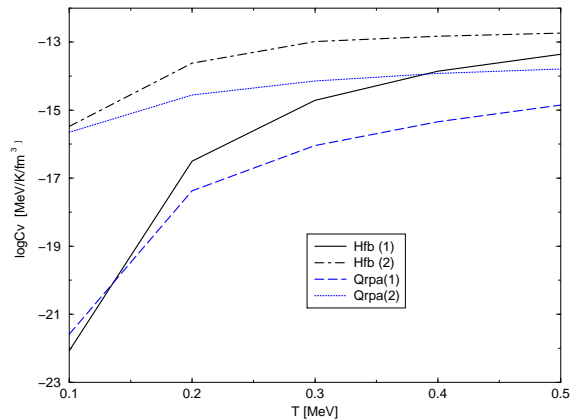}
\caption{ Specific heat in the cell $^{1800}$Sn. The dashed line corresponds
 to the collective modes and the full line to the HFB spectrum}
\end{center}
\end{figure}

\subsection{Cooling time of the inner crust}

The specific heat of the inner crust matter is an important quantity for cooling
time calculations. In this section we shall discuss the sensivity of the cooling
time on the specific heat calculated with the two scenarios for the nuclear
superfluidity introduced in section 3.1. The cooling process  we analyse here
corresponds to a fast cooling mechanism (e.g., induced by direct Urca reactions).
In this case the interior of the star cools much faster than the crust.
The cooling time is defined as the time needed for the cooling wave to traverse
the crust and to arrive at the surface of the star. According to numerical simulations
\cite{lattimer}, the cooling time is proportional to the square of the crust size and
to the specific heat. This result was used by Pizzochero et al \cite{pizzochero} for
estimating the cooling time in a simple model, which we have also employed in our
calculations \cite{cooling}. Thus, the crust is devided in shells of thickness $R_i$
for which the thermal difusivity $D_i$ could be considered as constant. Then the cooling
time is obtained by summing the contribution of each shell, i.e.,
\begin{equation}
t_{cool}=\frac{1}{\gamma } \sum_{i} \frac{R_{i}^2}{D(i,T)}  
\end{equation}
The thermal diffusivity is given by $D=\frac{\kappa}{C_{V}}$, where $\kappa$
is the thermal conductibility and $C_V$ is the specific heat.
The thermal conductivity is mainly determined by the electrons and in our
calculations we have used the values reported by Lattimer et al.~\cite{lattimer}. 
The specific heat $C_V$ has major contributions from the electrons, which
can be easily calculated,  and from the neutrons of the inner crust. 
As we have seen above, the specific heat of the neutrons depends strongly
on pairing correlations. In order to see if this dependence has observationally
consequences for the cooling time, we have performed two calculations, corresponding
to the strong and the weak pairing forces introduced in section 3.1. The calculations
show that the cooling time is increasing with more than $80\%$ if for calculating 
the pairing correlations we use a weak pairing force instead of a strong pairing force. 
This result
indicates that the nuclear superfluidity of the inner crust matter plays a crucial role 
for the cooling time calculation of neutron stars.

\section{Conclusions}

In this lecture we have shown how the HFB approach and the linear response theory
can be used to describe the pairing correlations in exotic nuclei
and in the inner crust of neutron stars. Thus, for the nuclei close to the 
drip lines we have discussed how one can incorporate in the two approaches mentioned above
the effects of the continuum coupling on pairing correlations. Then, using the same models,
we have analysed  what are the effects of pairing correlations on the specific heat and on the
cooling time of neutron stars. We thus found that the cooling time depends very strongly on the
intensity of pairing correlations in nuclear matter. Because the intensity of pairing correlations
in nuclear matter is still unclear, at present one can only estimate the limits in which the cooling
time of the inner crust can vary. Since the pairing correlations in nuclear matter and in nuclei are
in fact correlated, one hopes that these limits could be reduced by systematic studies of pairing
in  both infinite and finite nuclear systems.

\end{document}